\documentclass[prl,twocolumn,showkeys,preprintnumbers,amsmath,amssymb]{revtex4}
\usepackage{graphicx}% Include figure files
\usepackage{dcolumn}% Align table columns on decimal point
\usepackage{bm}% bold math

\def\be{\begin{equation}}
\def\ee{\end{equation}}
\def\e#1{\label{#1}\end{equation}}
\def\bea{\begin{eqnarray}}
\def\eea{\end{eqnarray}}
\def\ea#1{\label{#1}\end{eqnarray}}

\def\bem#1{\begin{mathletters}\label{#1}}
\def\eml{\end{mathletters}}

\def\ket#1{{|#1\rangle}}
\def\bra#1{{\langle#1|}}

\def\4#1{{\bf{#1}}}
\def\8#1{{\widetilde{#1}}}
\def\eqref#1{(\ref{#1})}

\begin{document}
\title{How to Maximize the Capacity of General Quantum Noisy Channels}

\author{Goren Gordon}
\email{goren.gordon@weizmann.ac.il}
\author{Gershon Kurizki}
\email{gershon.kurizki@weizmann.ac.il}
\affiliation{%
Department of Chemical Physics, Weizmann Institute of Science,
Rehovot 76100, Israel
}%

\date{\today}% It is always \today, today,

\begin{abstract}
A general quantum noisy channel is analyzed, wherein the transmitted qubits may experience symmetry-breaking decoherence, along with memory effects.
We find the optimal basis not to be fully entangled, but a combination of factorized and partially-entangled states in the presence of memory, asymmetry and the state-bias of the noise.
Capacity-maximization is shown to be achievable by combining temporal shaping of the transmitted qubits and optimal basis selection.
\end{abstract}

  \maketitle
%Introduction - channels - previous results
{\em Introduction}.
Communication of classical information, encoded into quantum states of the input ensemble is naturally classified according to the type of input ensemble in question \cite{hol98,ben02,bal04,mac02,bow04,kre05,gio05,mus08}.
One type is restricted to classical correlations between consecutive uses and can be prepared by adjusting the
quantum state of each particle sent through the channel. The other type of input ensembles consist of many entangled particles that are sent through the channel one by one. 
Both theory \cite{mac02,bow04,kre05,gio05} and experiment \cite{ben02,bal04} showed that whenever the noise affecting consecutive uses is correlated, i.e., has ``memory'', an entangled input ensemble can substantially enhance the classical capacity of a noisy depolarizing channel. 

%Introduction - a more general channel
In this Letter, we extend the theoretical analysis to a broader class of noisy channels, wherein the decoherence or noise may differ from qubit to qubit, here dubbed asymmetric (i.e. qubit symmetry-breaking),  and be state-biased, i.e., differ from one state to another, for each qubit, as well as exhibit memory effects, i.e., noise correlations of consecutive qubits.
We show that the optimal transmission basis is the one that lifts most degeneracies with respect to the decoherence/noise, meaning that all basis states have different decoherence rates.
Thus, while for a fully symmetric, unbiased memory channel, known as the depolarizing memory channel, the fully entangled basis is optimal as previously shown \cite{mac02,bow04,kre05,gio05}, we find that 
the combination of state-bias, asymmetry and memory of the noise results in a unique optimal basis that is {\em not fully entangled}.

The experimental scenario envisaged is a standard optical fiber with fluctuating birefringence,
which scrambles the polarization of photons.
Correlations of the birefringence act as memory, while its state-bias (preference of one polarization) can result in a final partial mixed state.
We show that dynamic control can be effected by temporal shaping of the qubit photon pulses, so as to take advantage of their unique optimal basis and thereby maximize the channel capacity.

%Protocol
{\em Protocol.} Alice transmits to Bob classical information
through a quantum noisy channel using two qubits as her information basis. 
This lossy channel has non-Markovian temporal memory that affects the second qubit after the first one has gone through.

Alice can encode the classical information in either a factorized, or (at least partially) entangled information basis. 
The qubit states are denoted here by $\ket{0,1}_{1,2}$, where the subscripts label the first or second qubit, respectively. 
We thus define the possible transmitted states as the following
orthonormal basis: 
\bea
&&\ket{\psi_1(m_\Phi)}=M_\Phi(\ket{00}+m_\Phi\ket{11}) \\ 
&&\ket{\psi_2(m_\Phi)}=M_\Phi(m_\Phi^*\ket{00}-\ket{11})\\
&&\ket{\psi_3(m_\Psi)}=M_\Psi(\ket{01}+m_\Psi\ket{10}) \\
&&\ket{\psi_4(m_\Psi)}=M_\Psi(m_\Psi^*\ket{01}-\ket{10})
\eea
Here the normalization $M_{\Phi,\Psi}=1/\sqrt{1+|m_{\Phi,\Psi}|^2}$ and $m_{\Phi,\Psi}$ measure the entanglement of the different basis states.
The first and second pairs of basis states, labeled by $m_{\Phi,\Psi}$, become the appropriate Bell-states in the fully entangled case, $m_{\Phi,\Psi}=1$ and factorized states for $m_{\Phi,\Psi}=0$.

%Measure
The channel input is assumed to be the pure-state basis $\psi_x$, where $x=1\ldots 4$, while its output is, due to its interaction with the environment, the mixed-state $\rho_x$.
As a measure of the performance, we shall use the Holevo channel capacity \cite{hol98}:
\bea
\label{channel-capacity-def}
&&{\rm C}(t)=\max_{p_x,m_\Phi,m_\Psi}\chi(\{p_x,\4\rho_x(t)\})\\
\label{chi-def}
&&\chi(\{p_x,\4\rho_x\})=S(\langle\4\rho\rangle)-\sum_x
p_xS(\4\rho_x)\\
&&\label{s-def}
S(\4\rho) = -{\rm Tr}(\4\rho\log\4\rho)
\eea
with $\langle\4\rho\rangle=\sum_x p_x\4\rho_x$, $p_x$ is the
probability of sending $\ket{\psi_x}$, and $S(\4\rho)$ is the
Von-Neumann entropy. 
The goal of the present work is to find the basis that attains the maximal channel capacity as a function of the channel decoherence parameters and dynamically control qubits to exploit it.

%Decoherence parameters
{\em Channel decoherence parameters.}
The crucial point is that the channel capacity depends non-linearly on the decoherence of the transmitted basis. 
This means that increasing the decoherence of one basis state while decreasing that of another results in {\em increased} channel capacity, compared to a basis wherein all states experience the same decoherence.
Hence, the optimal basis {\em lifts the degeneracies} with respect to the decoherence.
 
The decoherence of the transmitted qubits may (completely generally) be characterized by the following parameters: 
(i) For a single qubit, the $\ket{0}_{jj}\bra{0}\rightarrow\ket{1}_{jj}\bra{1}$ and $\ket{1}_{jj}\bra{1}\rightarrow\ket{0}_{jj}\bra{0}$ transitions may have $R_{jj}^0$ and $R_{jj}^1$ rates, respectively. 
{\em Decoherence-rate magnitude} is defined by $\nu^{0(1)}=R_{11}^{0(1)}+R_{22}^{0(1)}$, where for the case of $\nu^{0(1)}=0$, we have no decoherence, i.e., completely preserve information and entanglement. 
(ii) {\em Decoherence state-bias} is defined as $\alpha=1-\nu^0/\nu^1$ and measures whether one of the qubit's states is ``preferred'' by the channel. 
(iii) Decoherence may discriminate between consecutive qubits, i.e., $R_{jj}^{0,1}\neq R_{kk}^{0,1}$ for $k\neq j$.
This {\em decoherence rate asymmetry} is defined as $\zeta=|R_{11}^{1}-R_{22}^{1}|/\nu^1$.
Then $\zeta=0$ is the fully symmetric case where both qubits experience the same decoherence. 
The case of full asymmetry, $\zeta=1$, means that one qubit is fully protected at the expense of the other qubit, which experiences increased decoherence.
(iv) The decoherence of qubit-pair transmission, described by the $\ket{0}_j\ket{1}_k\rightarrow\ket{1}_j\ket{0}_k$ transition, may exhibit memory embodied in the $R_{jk}^{0,1}$ transition rates: the noise of one qubit affects the noise of the subsequent qubit.
This {\em memory, or cross-decoherence}, parameter is defined as $\mu=|R_{12}^{1}+R_{21}^{1}|/\nu^1$, where
$\mu=1$ denotes maximal memory, and $\mu=0$ denotes completely uncorrelated noise. 

In the case of full asymmetry, $\zeta=1$, the entangled singlet and triplet, $\ket{\psi_{3,4}(m_\Psi=1)}$, are degenerate, i.e. experience the same decoherence.
By contrast, in the factorized basis, the states $\ket{\psi_{3,4}(m_\Psi=0)}$ and $\ket{\psi_{1,2}(m_\Phi=0)}$, experience different decoherence, since one basis state does not decohere while the other decoheres rapidly.
Thus, the factorized basis {\em lifts this degeneracy} of a fully asymmetric channel and is then optimal.

Now, consider a full memory channel, $\mu=1$. 
It is known that the singlet, $\ket{\psi_4(m_\Psi=1)}$, does not experience decoherence in this channel, while the triplet, $\ket{\psi_3(m_\Psi=1)}$ experiences increased decoherence \cite{mac02,gio05}. 
By contrast, the factorized pair of basis states, namely $\ket{\psi_{3,4}(m_\Psi=0)}$, experience the same decoherence in the full memory channel.
Hence, the pair of entangled states $\ket{\psi_{3,4}(m_\Psi=1)}$ {\em lifts the degeneracy} with respect to the memory of the channel and is thus optimal when $\mu=1$.
The other pair of states, $\ket{\psi_{1,2}(m_\Phi)}$ is however degenerate for all $m_\Phi$, i.e. is not influenced by the memory.
We therefore consider other decoherence attributes that affect the $\ket{\psi_{1,2}(m_\Phi=1)}$ pair of entangled states.

In a fully state-biased channel, $\alpha=1$, the factorized $\ket{\psi_1(m=0)}=\ket{00}$ state does not experience decoherence, while the factorized $\ket{\psi_2(m_\Phi=0)}=\ket{11}$ state experiences increased decoherence.
On the other hand, the fully entangled pair of states $\ket{\psi_{1,2}(m_\Phi=1)}$ are degenerate with respect to the bias, i.e., experience identical decoherence. 
Hence, the factorized basis pair {\em lifts the degeneracy} with respect to bias.
On the other hand, bias does not affect the other pair of basis states, $\ket{\psi_{3,4}(m_\Psi)}$, which is degenerate for all $m_\Psi$.

We then reach a remarkable conclusion: the fully entangled (Bell) basis is not always optimal.
If asymmetry is present, the fully factorized basis becomes optimal.
If bias is present in a memory channel, a unique basis becomes optimal, namely the one composed of the fully entangled pair $\ket{\psi_{3,4}(m_\Psi=1)}$ and the fully factorized pair $\ket{\psi_{1,2}(m_\Phi=0)}$.
Only for an unbiased, symmetric depolarizing channel, is the fully entangled basis optimal.

To study the transition from one optimal basis to another, one must analyze the entire range of memory, $\mu$, asymmetry, $\zeta$ and state-bias, $\alpha$, where $0\ge\left\{\zeta,\mu,\alpha\right\}\ge1$, by scanning through the parameter space using the inequality $\zeta^2+\mu^2\le 1$.

{\em Channel capacity analysis.}
We compare the following extreme limits: 
(a) symmetric, state-biased memory channel, $\zeta=0$, $\alpha=1$;
(b) asymmetric, state-biased memoryless channel, $\mu=0$, $\alpha=1$;
(c) symmetric, unbiased memory channel, $\zeta=\alpha=0$ and;
(d) asymmetric, unbiased memoryless channel, $\mu=\alpha=0$.

\begin{figure}[ht]
\centering\includegraphics[width=8cm]{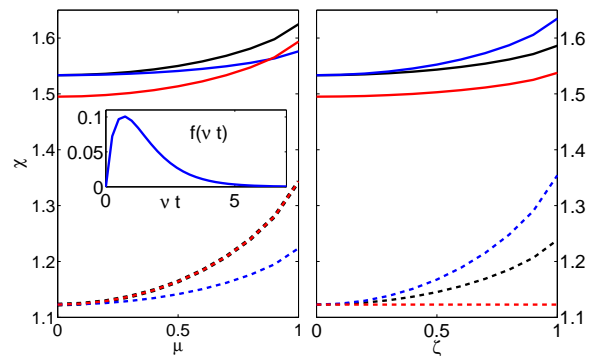}
 \caption{Channel capacity as a function of channel parameter.
 (a) Dependence of biased (solid) and unbiased (dashed) channel capacity on memory, $\mu$, for combined (black), factorized (blue) and Bell (red) basis. Inset: $f(\nu t)$.
 (b) Dependence of channel capacity on asymmetry, $\zeta$.
 Here $\nu t=0.1$.}
 \protect\label{figure-chi_p}
\end{figure}

While {\em analytical results} have been obtained for all these limits, they are extremely cumbersome. 
One can approximate these analytic results, for a given decoherence rate magnitude, $\nu$, via the following relations for the channel capacities in the different cases:
\bea
&&\chi_b^{\rm fac}(p) = \chi_a^{\rm ent}(p) + \epsilon_{ba}(p) + f(\nu t)\\
&&\chi_a^{\rm fac}(p) = \chi_b^{\rm ent}(p) + \epsilon_{ab}(p) + f(\nu t)\\
&&\chi_d^{\rm fac}(p) = \chi_c^{\rm com}(p) + \epsilon_{dc}(p)\\
&&\chi_c^{\rm fac}(p) = \chi_d^{\rm com}(p) + \epsilon_{cd}(p)
\eea
where the $a,b,c$ and $d$ subscripts label the limits listed above and the superscripts label the factorized, entangled and combined bases. 
Here, $\chi_{a,c}(p)=\chi_{a,c}(\mu)$ and $\chi_{b,d}(p)=\chi_{b,d}(\zeta)$, all $\epsilon(p)\ll1$ and $f(\nu t)\ge 0$ and is given by
\bea
f(\nu t) &=& 0.5-\Big[(\ln(d_1d_2/d_3)-2\nu t)+(r\ln(d_2/d_1)+ \nonumber\\&&\ln(d_3^2/(d_1d_2)))/w+\ln(d_1d_2/d_3)/w^2\Big]/\ln(16)\\
w&=&e^{\nu t},\quad r=\sqrt{1+d_3},\quad d_3=(1-w)^2,\\
d_1&=&d_3+w(1-r),\quad d_2=d_3+w(1+r).
\eea

As can be seen, (Fig.~\ref{figure-chi_p}) our optimal basis analysis is confirmed by these results.
The fully entangled basis is {\em always} suboptimal for the state-biased memory channel, due to the degeneracy of the $\ket{\psi_{1,2}(m_\Phi=1)}$ basis pair, whereas the suggested combined basis is optimal under bias.
The results also confirm that the factorized basis is optimal for the asymmetric channel.
 
For the state-biased case, the dependence of the entangled-basis capacity on memory is similar to that of the factorized basis on asymmetry.
For the unbiased case, the effects of memory on the combined basis, i.e. the combination of a factorized pair and a fully entangled pair, are similar to the effects of asymmetry on the factorized basis.
This corroborates our claim that only {\em degeneracy-lifting} affects the channel capacity, rather than specific details of decoherence mechanisms.

\begin{figure}[htb]
\centering\includegraphics[width=8.5cm]{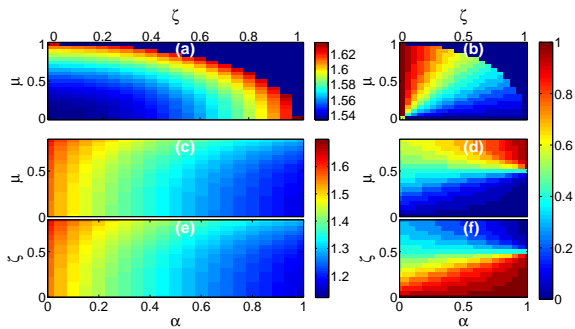}
\protect\caption{Channel capacity as a function of: 
(a) memory ($\mu$) and asymmetry ($\zeta$) for a fully state-biased noise ($\alpha=1$);
(c) memory ($\mu$) and bias ($\alpha$) for partial noise asymmetry ($\zeta=0.5$);
(e) asymmetry ($\zeta$) and state-bias ($\alpha$) for partial memory ($\mu=0.5$).
Optimal transmission basis is given by a combination of a factorized pair of states $\ket{\Psi_{1,2}(m_\Phi=0)}$ and the {\em partially entangled} pair  of states $\ket{\Psi_{3,4}(m_\Psi)}$; (b),(d) and (f) show $m_\Psi$ for decoherence rate parameters given in (a), (c) and (e), respectively.
Here, $\nu t=0.2$.
}\label{figure-channel}
\end{figure}

Since in realistic scenarios asymmetry, memory and state-bias are generally not in the extreme limits, it is imperative to analyze intermediate values of these parameters (between $0$ and $1$) and their implications on the optimal basis (Fig.~\ref{figure-channel}).
When two competing decoherence mechanisms that cause degeneracy-lifting effects are present, e.g. memory and asymmetry, the optimal basis is composed of the factorized pair, $\ket{\Psi_{1,2}(m_\Phi=0)}$, and a {\em partially entangled} pair, $\ket{\Psi_{3,4}(m_\Psi\neq0)}$ (Fig.~\ref{figure-channel}(b,d,f)).
The latter pair of states varies continuously from the factorized form, $m_\Psi=0$, when asymmetry and/or bias dominate, to the fully entangled one, $m_\Psi=1$ when memory dominates.
Thus, the relative strength of such competing decoherence mechanisms determines the degeneracy-lifting partial entanglement of the optimal basis.

{\em Active dynamic control.} 
This gives rise to the following questions: 
can decoherence parameters be controlled? 
Namely, can one {\em actively change} the asymmetry and memory of a given channel and thus optimize not only the transmitted basis, but also the channel capacity? 
To answer these questions we have developed a dynamic model of the channel, which is comprised of $N=2$ qubits that are weakly coupled to a thermal bath, composed of harmonic oscillators, with corresponding energies
$\hbar\omega_\lambda$.
The {\em temporal shape} of the qubits injected into the channel, denoted by $\epsilon_j(t)$, modulates the coupling to the thermal bath.
The total Hamiltonian is given by (henceforth, $\hbar=1$):
\bea
\label{total-Hamiltonian}
&&H(t)=H_S + H_B + H_I(t),\quad H_B=\sum_\lambda \omega_\lambda a_\lambda^\dagger a_\lambda\\ 
\label{system-Hamiltonian}
&&H_S=\sum_{j=1}^N \omega_{aj}\ket{1}_{jj}\bra{1}\bigotimes_{j'\neq j} I_{j'},\\
\label{int-Hamiltonian}
&&H_I(t)=\sum_{j=1}^N\sum_\lambda\sigma_{x,j}(\epsilon_j(t)\kappa_{\lambda,j}a_{\lambda,j}+H.c.)\bigotimes_{j'\neq j}I_{j'}
\eea
%\label{bath-Hamiltonian}
Here the possible non-degeneracy of the states leading to bias is denoted by frequencies $\omega_{aj}$,
where $\sigma_{x,j}=\ket{1}_{jj}\bra{0}+\ket{0}_{jj}\bra{1}$ is
the X-Pauli matrix of qubit $j$, $\4I$ is the identity matrix, $H.c.$ is Hermitian conjugate and
$\kappa_{\lambda,j}$ is the off-diagonal coupling coefficient of
qubit $j$ to the bath oscillator $\lambda$. $\epsilon_j(t)=\tilde\epsilon_j(t)e^{i\phi_j(t)}$, where $\tilde\epsilon_j(t)$ and $\phi_j(t)$ are the time-dependent amplitude and phase of the pulse. 
Note that we did {\em not invoke the rotating-wave approximation} which may fail for fast modulations \cite{kof04}.
%Our control is supplementary to the active polarization control that has been successfully implemented \cite{vitali}.

\begin{figure}[ht]
\centering\includegraphics[width=8cm]{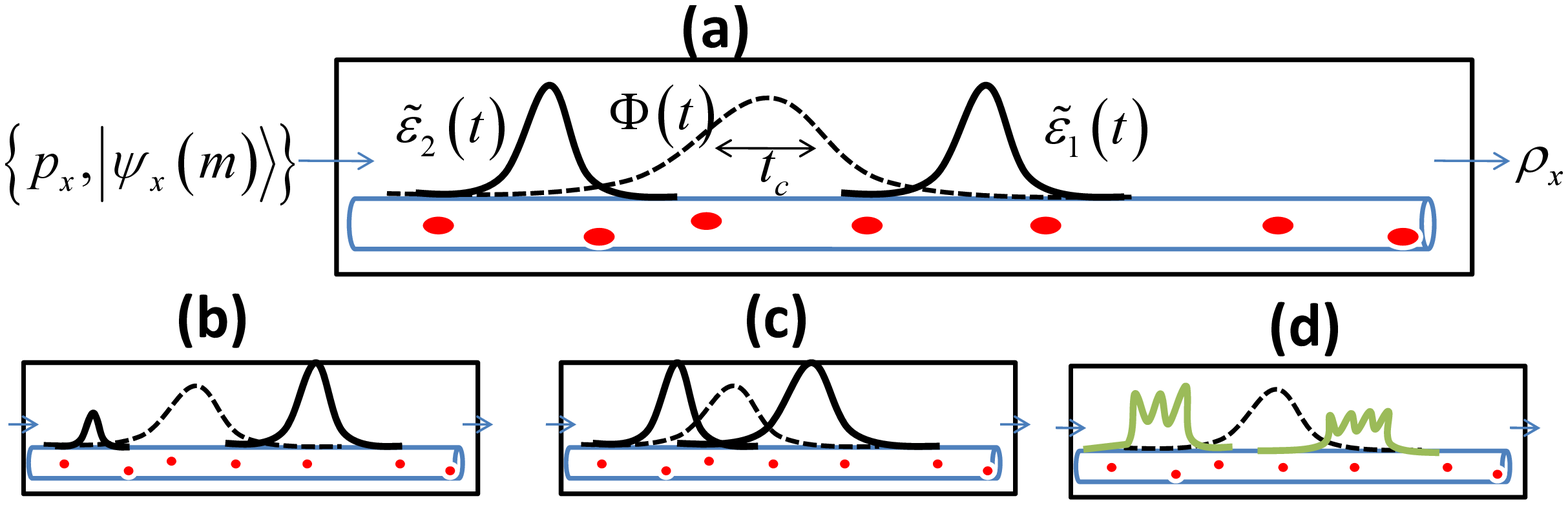}
 \caption{Schematic diagram of the channel.
 (a) Two consecutive temporally shaped qubits $\epsilon_j(t)$ (solid) and the system-bath correlation function $\Phi(t)$ (dashed), with a typical correlation time $t_c$.
 The bath (noise) correlations are represented by the red dots.
 (b) Large asymmetry ($\zeta$), i.e. different qubit temporal shapes.
 (c) Large memory (cross-decoherence, $\mu$), i.e. larger temporal overlap mediated by the correlation function.
 (d) Large state-bias, i.e. different chirp of $\ket{1}$ and $\ket{0}$ (green) in each qubit.
 }
 \protect\label{figure-schematic}
\end{figure}

It can be shown \cite{gor06b,gor08PRL} that the non-Markov dynamics of the reduced system density matrix is primarily determined by the following time-dependence of the qubit-pairs decoherence rates, $R_{jk}(t)$ of qubit pairs
\be
\label{R-def}
R_{jk}(t)=2{\rm Re}\int_0^tdt_1 \Phi_{T,jk}(t-t_1)\epsilon_j(t)\epsilon^*_{k}(t_1) e^{i(\omega_{aj}t-\omega_{ak}t_1)}
\ee
where $\Phi_{T,jk}(t)$ is the temperature-dependent channel correlation (response) function.
Since Alice uses a single channel that is identical for the two qubits, it has $\Phi_{T,jj'}(t)=\Phi_T(t)$, and is characterized by a typical correlation time $t_c$ over which non-Markovian dynamics occurs (Fig.~\ref{figure-schematic}(a)).

Equation~\eqref{R-def} shows that the qubits' temporal shape can effectively control the decoherence parameters.
Asymmetry can be modified by having a {\em different temporal shape} for each qubit, $\epsilon_j(t)\neq\epsilon_k(t)$, Fig.~\ref{figure-schematic}(b), while memory can be decreased by increasing the delay or changing the phase variation of the pulse, $\phi_j(t)$ \cite{gor06b} (Fig.~\ref{figure-schematic}(c)).

Bias is associated with $\omega_{aj}\neq0$, the energy difference between the two qubit levels.
An unbiased channel is equivalent to an infinite temperature bath or to a degenerate qubit.
By introducing high frequency components (chirp) in the photonic qubit phase $\phi_j(t)$ and/or controlling $\omega_{aj}$ one can modify the amount of state-bias \cite{wei90} (Fig.~\ref{figure-schematic}(d)).

{\em Experimental setup.}
A model of a noisy quantum channel \cite{bal04} was motivated by fluctuating birefringence of a standard optical fiber,
which scrambles the polarization of an input light pulse to a completely mixed state, without bias, $\alpha=0$. 
Since the characteristic time scale of birefringence fluctuations is usually much longer than the temporal separation between consecutive light pulses full memory prevails, $\mu=1$. 
The spectral dispersion of the fluctuations makes them sensitive to the input pulse shape.

One can thus encode the classical information in the polarization of the two pulses, each pulse containing exactly one photon. 
The advantage of employing entangled states in the above scenario becomes obvious when we recall that the
singlet polarization state of two photons remains invariant under correlated depolarization \cite{kwi00}, and therefore can be unambiguously discriminated against the orthogonal triplet subspace \cite{bar03}.
On the other hand, the difference in the consecutive pulses' shapes \cite{wei90} may control their asymmetry and state-bias of their decoherence, rendering the Bell basis suboptimal.

%conclusions
{\em Conclusions.} 
We have analyzed the properties of a completely general quantum noisy channel for classical communication. 
To this end, we have extended previous works \cite{mac02} in search of the optimal basis for the set of the decoherence parameters: magnitude, asymmetry, memory and state-bias. 
A control scheme of the decoherence rates has also been proposed, by shaping of individual qubit pulses sent through the channel.

We have shown that the optimal basis is the one that {\em lifts the degeneracies} with respect to these decoherence characteristics.
Thus, the singlet-triplet basis pair lifts the degeneracy of a memory channel, while the factorized basis pair lifts the degeneracy of an asymmetric channel.
In particular, we have shown that optimal two-photon encoding depends on the state-bias of the channel.
Thus, optimal encoding requires to factorize the triplet subspace, and not use the entangled basis.
Accordingly, we have demonstrated that the most general optimal basis is composed of a factorized pair and a {\em partially entangled} pair whose entanglement depends on the relative strength of competing decoherence mechanisms.

Knowing, or better still, controlling the decoherence parameters by photon pulse shaping, that may be optimized (in particular, chirped) under pulse energy constraints \cite{gor08PRL}, and thus adapted to an optimal basis choice for these parameters, suggests a practical approach to maximizing the channel capacity for information encoded in photon polarization.

We acknowledge the support of ISF, GIF and EC (MIDAS and SCALA Projects).

%\bibliography{Bibliography}   %>>>> bibliography data in report.bib
%\bibliographystyle{apsrev}   %>>>> makes bibtex use spiebib.bst

%mac02,mac04,ars06
%gor06b,gor08EPL
%wei90,fru08
\end{document}